\documentclass[a4paper,fleqn,usenatbib]{mnras}

\pdfoutput=1

\usepackage{newtxtext,newtxmath}

\usepackage[T1]{fontenc}
\usepackage{ae,aecompl}

\usepackage{graphicx}	
\usepackage{amsmath}	

\usepackage{newtxtext,newtxmath}

\usepackage{xcolor}
\pagecolor{white}

\title[ANN Classification of 4FGL Sources]{Artificial Neural Network Classification of 4FGL Sources}
\author[S. Germani et al.]{
S. Germani,$^{1}$\thanks{E-mail: stefano.germani@unipg.it}
G. Tosti,$^{1}$
P. Lubrano,$^{2}$
S. Cutini,$^{2}$
I. Mereu,$^{2}$
A. Berretta$^{1}$\\
$^{1}$Dipartimento di Fisica e Geologia, Univ. degli Studi di Perugia, Via A. Pascoli snc, I-06123 Perugia, Italy\\
$^{2}$INFN – Istituto Nazionale di Fisica Nucleare Sez. Perugia, Via A. Pascoli snc, I-06123 Perugia, Italy\\
}

\date{Accepted 2021 June 15. Received 2021 June 14; in original form 2021 May 13.}

\pubyear{2021}

\begin{document}
\maketitle

\begin{abstract}
The \emph{Fermi}-LAT DR1 and DR2 4FGL catalogues feature more than 5000 gamma-ray sources of which about one fourth are not associated with already known objects, and approximately one third are associated with blazars of uncertain nature. We perform a three-category classification of the 4FGL DR1 and DR2 sources independently, using an ensemble of Artificial Neural Networks (ANNs) to characterise them based on the likelihood of being a Pulsar (PSR), a BL Lac type blazar (BLL) or a Flat Spectrum Radio Quasar (FSRQ). We identify candidate PSR, BLL and FSRQ among the unassociated sources with approximate equipartition among the three categories and select ten classification outliers as potentially interesting for follow up studies.
\end{abstract}




\section{Introduction}


The identification of sources with counterparts at other wavelengths has been a significant issue in gamma-ray astrophysics since its beginnings due to the limited angular resolution of instrumentation. 

The \emph{Fermi}-LAT~\citep[][]{2009ApJ...697.1071A} has revolutionised the field, increasing the number of known gamma-ray sources by more than an order of magnitude and opening a new energy range window between the tens and hundreds of GeV. The \emph{Fermi}-LAT collaboration has published several catalogues, named 1 to 4FGL~\citep{1FGL,2FGL,3FGL,4FGL}, in succession. The 4FGL catalogue, after a first release corresponding to 8 years of data (4FGL-DR1\footnote{\url{https://fermi.gsfc.nasa.gov/ssc/data/access/lat/8yr_catalog/}}), has been updated with an incremental 10 year version named 4FGL-DR2\footnote{\url{https://fermi.gsfc.nasa.gov/ssc/data/access/lat/10yr_catalog/}} \citep{4FGLDR2}.
As highlighted in Table~\ref{tab:xFGL}\footnote{The xFGL catalogues distinguish between firmly identified, and associated sources indicated with capital and lower case class names respectively. We consider both identified and associated sources together and use class names with capital letters only.}, the number of detected gamma-ray sources and source types has steadily increased over time. Despite the astrophysical community's efforts to match many previously unidentified sources with possible counterparts through intensive multi-frequency campaigns, each new catalogue shows a substantial fraction of sources without a plausible association (UID). The majority of \emph{Fermi}-LAT sources are Pulsars (PSR) and Blazars (BLZ), the latter are divided between the BL Lac type sources (BLL), Flat Spectrum Radio Quasars (FSRQ) and Blazars of unknown type (BCU): BLZ without a clear optical spectral identification.
\begin{table}
	\centering
	\caption{Evolution of \emph{Fermi}-LAT xFGL catalogues.}
	\label{tab:xFGL}
	\begin{tabular}{lccrrc} 
		\hline
		Catalogue & Source Types & Sources & UID  (\%)  &  BCU  (\%) \\
		\hline
		 1FGL        &  11 & 1451 & 630  (43) & 92   (6)  \\
		 2FGL        &  13 & 1873 & 575  (31) & 257  (14) \\
		 3FGL        &  18 & 3033 & 1010 (33) & 568  (19) \\
		 4FGL-DR1    &  19 & 5064 & 1336 (26) & 1310 (26) \\
		 4FGL-DR2    &  19 & 5788 & 1679 (29) & 1515 (26) \\				  
		\hline
	\end{tabular}
\end{table}

Starting from~\citet{2012ApJ...753...83A} using the 1FGL catalogue, several studies have used Machine Learning (ML) to characterise the unassociated sources based on their gamma-ray properties~\citep{2012MNRAS.424.2832L, 2014ApJ...782...41D, 2017MNRAS.470.1291S,2016ApJ...820....8S}.
Other works aimed at an ML-based classification of BCU~\citep{2013MNRAS.428..220H, 2017ApJ...838...34Y, 2019ApJ...872..189K}, in particular~\citet{2016MNRAS.462.3180C}, introduced the use of the empirical cumulative distribution function of the Light Curves (LC), sorted in flux, as part of the input to an Artificial Neural Network for 3FGL sources, the same technique was also used in later works on BCU from the 4FGL DR1 catalogue~\citep{2019MNRAS.490.4770K,2020MNRAS.493.1926K}. The study from~\citet{2016ApJ...825...69M}, together with identifying potential pulsar candidates within the 3FGL  unassociated sources, searched for peculiar sources with the possible interpretation of dark matter subhalos. In all cases, a two-category classification was performed (PSR-BLZ, BLL-FSRQ).

In this work, we present an alternative study to characterise unassociated sources in the DR1 and DR2 4FGL catalogues using an ANN as a three-category classifier to identify candidate PSR, BLL and FSRQ sources. Furthermore, we select those whose classification results mark them as dissimilar from other known source classes to identify them as interesting for follow-up studies or multiwavelength campaigns. In Section~\ref{sec:method} we describe our method featuring the chosen approach, the developed ANN model and the training strategy. The classification results, including validation checks and sources selected as special cases, are discussed in Section~\ref{sec:results}, and our conclusions are outlined in Section~\ref{sec:conclusions}.

\section{Method}
\label{sec:method}

\subsection{Approach}
\label{sec:approach}

The 4FGL DR1 and DR2 catalogues offer similar gamma-ray characteristics as the other xFGL general catalogues. The principal difference is the release of additional light curve information, in fact for the DR1 catalogue light curves are available with both two month and one-year-long time bins (48 and 8 flux data points respectively for 8 years of integration), while the DR2 catalogue offers just 10 one-year long time bins for the light curves. We chose to develop two similar, but independent, ANNs to leverage both catalogues information: the more finely sampled light curves from the DR1 and the longer integration time corresponding to the DR2. This also lends the possibility to compare and cross-correlate results.

We perform a multi-category classification based on the three most populous source types corresponding to well-identified objects with known characteristics: BLL, FSRQ and PSR. The number of sources belonging to the main classes is shown in Table~\ref{tab:4FGL}, where all the source types with fewer representatives are collectively indicated with the OTH class.  
\begin{table}
	\centering
	\caption{Census of the 4FGL source categories for DR1 and DR2. The column OTH combines all the sources other than BLL, FSRQ, PSR, BCU, and UID.}
	\label{tab:4FGL}
	\begin{tabular}{crrrrrr} 
		\hline
		DR & BLL  & FSRQ &  PSR  &  BCU  & UID & OTH \\
		\hline
		 1   & 1131  &   694  &  239   & 1310   &  1336  &  355  \\
		 2   & 1190   &  730  &  259   &  1515  &  1679  &  415  \\		  
		\hline
	\end{tabular}
\end{table}

\begin{figure}
	\includegraphics[width=\columnwidth]{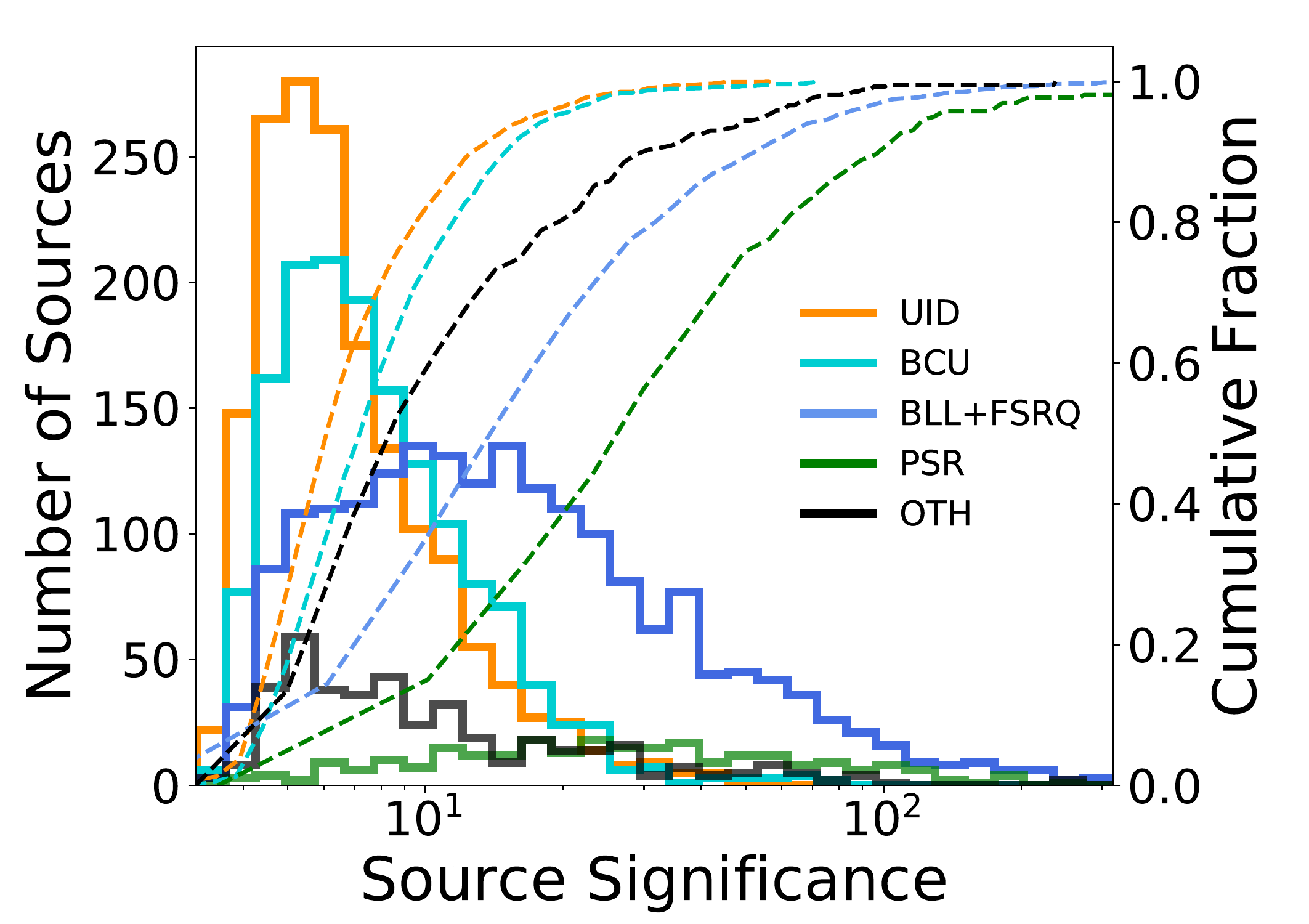}
  \caption{The detection significance ($Signif\_Avg$) of different source classes in the 4FGL DR2 catalogue.  Solid lines show the number of sources per significance interval, dashed lines the corresponding cumulative fraction.}
  \label{fig:signif_all}
 \end{figure}

The source detection significance, corresponding to the $Signif\_Avg$ column in the 4FGL catalogue, is not uniform among the different classes, as shown in Figure~\ref{fig:signif_all} (DR2 only). Pulsars are the sources with the highest average significance, followed by Blazars (BLL+FSRQ). UID and BCU are limited to low values of significance, with the distribution for sources belonging to all the other classes (OTH) peaking at low significance, but showing a relatively long tail towards large values.

To search for peculiar unassociated sources, we choose to only select ANN input quantities based on astrophysical motivations, avoiding automated variable ranking which is likely to select the best quantities only for the classification of the already known source types.

\subsection{Artificial Neural Network Model}
\label{sec:model}

The three classes chosen for classification are characterised by different features in their gamma-ray spectrum and light curves~\citep{4FGL, 4LAC, LATBriightBLZ,2014MNRAS.441.3591H}. Pulsars have characteristically curved spectra with pulsation periods on the order of seconds or smaller, resulting in a stable flux when integrated over the time scale of months or years. Blazars are typically variable sources with a power-law spectrum; however, some can show a break or a curvature. BLLs tend to have a harder spectrum and tend to be less variable in comparison with FSRQs.

The variables used for classification are directly taken or derived from those available in the 4FGL catalogue, with all related to the sources' spectral and time variability features. The catalogues provide the flux at seven different energy bands; we use six hardness ratios (HR) defined as:
\begin{equation}
   HR_{i} = \frac{F_{i-1} - F_{i}}{F_{i-1} + F_{i}}
	\label{eq:hr}
\end{equation}
where $F_{i}$ is the flux in the $i^{th}$ energy band corresponding to the \textit{Flux\_Band} column.
As previously mentioned, we used the two-month (\textit{Flux2\_History}) and one-year-long (\textit{Flux\_History}) time interval light curves for the DR1 and DR2 respectively. We use both the unmodified LC and the flux sorted Light Curves (LCs). Each time interval flux is divided by the average source flux over the whole time-span covered by the catalogue (\textit{Flux1000}). We set the value to zero for both the energy bands and light curve fluxes if the measurement resulted in an upper limit.
To summarise the variables derived from the spectral band and the light curves are:
\begin{description} 
\item the hardness ratios ($HR_{i}$);
\item the light curve (LC);
\item the flux sorted light curve (LCs).
\end{description} 
Other variables, collectively called \textit{IDX} henceforth, directly extracted  from the catalogue and used as input to the ANN are:
\begin{description} 
\item the logarithm of the variability index (\textit{Variability\_Index});
\item the fractional variability (\textit{Frac\_Variability});  
\item the spectral index from the power law fit (\textit{PL\_index});
\item the $\beta$ parameter from the log parabola fit (\textit{LP\_beta});
\item the curvature significance from the log parabola fit (\textit{LP\_SigCurv}).
\end{description}

\begin{figure}
	\includegraphics[width=\columnwidth]{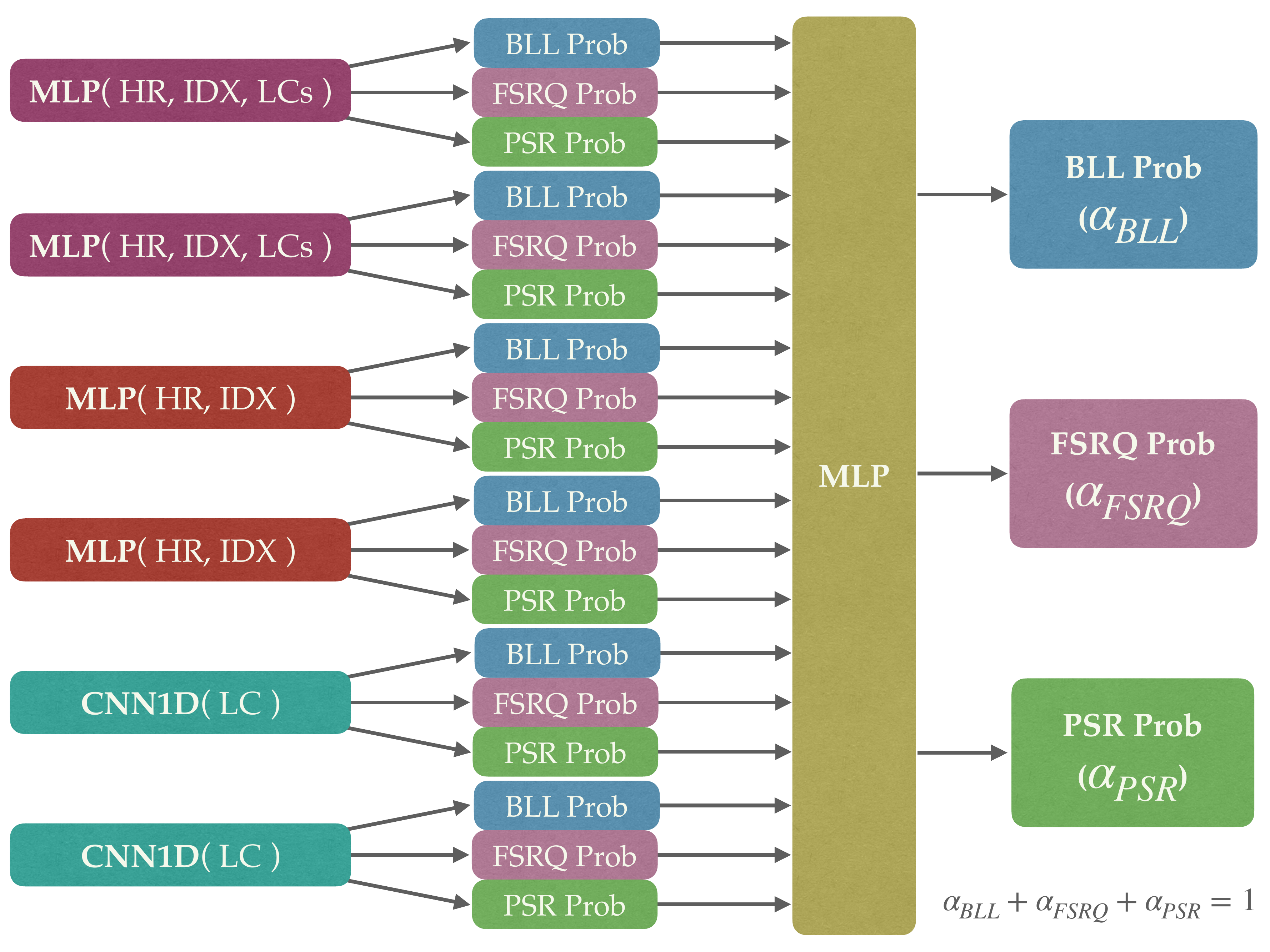}
    \caption{Artificial Neural Network stacked model diagram. The input and output variables are described in greater detail within the text.}
  \label{fig:stacked_model}
\end{figure}

The chosen approach is to develop and train a set of ANNs using different architectures and techniques which are then combined into a stacked ensemble (Figure~\ref{fig:stacked_model}) where the output from each network is fed as input to a dedicated Multi-Layer Perceptron (MLP)~\citep[][]{Rosenblatt1957} which provides the final classification estimating the probability to be a BLL ($\alpha_{BLL}$), a FSRQ ($\alpha_{FSRQ}$) or a Pulsar ($\alpha_{PSR}$) where $\alpha_{BLL} + \alpha_{FSRQ} + \alpha_{PSR} = 1$.
As shown in Figure~\ref{fig:stacked_model}, as individual models, we use:
\begin{description} 
\item two different implementations of an MLP with the HR, IDX and sorted LC variables as input;
\item two different implementations of an MLP with the HR and IDX variables as input;
\item two different implementations of a Unidimensional Convolutional Neural Network  (CNN1D;~\citealt{DeepLearning}) with only the LC as input;
\end{description}

The ANN results may depend, to some extent, on the starting values of the large number of network parameters involved, which are usually stochastically assigned. Therefore, the classification outcome is not wholly reproducible if the random seeds are changed. Additionally, ANNs can perform differently between the data sample used for training and other independent testing samples. Preliminary studies, performed to find the optimal model, show that the stacked models' classification results are more stable compared to individual models and that the difference in performance between the training and testing samples is reduced.

The software frameworks used for this study are the \textit{Python}\footnote{\url{https://www.python.org}} based \textit{Keras}\footnote{\url{https://keras.io}} Application Programming Interface \citep{KerasAPI2015} and the \textit{TensorFlow}\footnote{\url{https://www.tensorflow.org}} platform \citep{TensorFlowWP}.

\subsection{Training}
\label{sec:training}

The data sample used for training and testing of the ANNs includes only the 4FGL catalogue sources with a primary association (\emph{CLASS1} column) to BLL, FSRQ and PSR. We used a randomly chosen training sample corresponding to 77\% of the selected sources, the remaining fraction of the sample (33\%) is used to validate the classification results.

The training of the stacked ensemble involved two main steps:
\begin{enumerate}
\item each individual model was trained and optimised to reach the best performance while avoiding overtraining;
\item to find the optimal stacked ensemble weighting procedure, the weights of each individual trained model were then fixed, and only the output combining MLP was trained to produce the final classification result.
\end{enumerate}
Both the single and ensemble models have been trained independently for DR1 and DR2 catalogues.

Even though the samples for the three classification categories were slightly imbalanced (see Table~\ref{tab:4FGL}), assigning counterbalancing weights to the different categories during the training did not produce noticeable differences in performance; therefore, we choose not to use them. 

\section{Classification Results}
\label{sec:results} 

The trained ANN models are used to characterise the likelihood of being a BLL, a FSRQ or a PSR for all the sources in the DR1 and DR2 catalogues, independently of their class, even if the source does not belong to one of the three classification categories. For analysis purposes, the sources are assigned to the category with the highest classification probability ($\alpha$) without applying any minimum threshold.

The full table with the classification results is available online, here we present a sample extract for DR1 and DR2 in Table~\ref{tab:dr1_table} and Table~\ref{tab:dr2_table} respectively.

\begin{table*}
	\centering
	\caption{Extract of the 4FGL DR1 (8y) classification results. The full table is available online.}
	\label{tab:dr1_table}
	\begin{tabular}{ccccccccc} 
	\hline
Source Name              &   Class   &      L     &      B     &  $\alpha_{BLL}$ &  $\alpha_{FSRQ}$ &  $\alpha_{PSR}$&  $\alpha_{BLL}+\alpha_{FSRQ}$ & Classification \\
\hline
4FGL J0000.3-7355        &           &      307.71&      -42.73&       0.971&       0.025&       0.004&       0.996&     BLL \\
4FGL J0001.2+4741        & bcu       &      114.25&      -14.34&       0.753&       0.224&       0.023&       0.977&     BLL \\
4FGL J0001.2-0747        & bll       &       89.03&      -67.31&       0.960&       0.034&       0.006&       0.994&     BLL \\
4FGL J0001.5+2113        & fsrq      &      107.65&      -40.17&       0.030&       0.948&       0.021&       0.979&    FSRQ \\
4FGL J0001.6-4156        & bcu       &      334.23&      -72.03&       0.981&       0.016&       0.003&       0.997&     BLL \\
\hline
	\end{tabular}
\end{table*}

\begin{table*}
	\centering
	\caption{Extract of the 4FGL DR2 (10y) classification results. The full table is available online.}
	\label{tab:dr2_table}
	\begin{tabular}{ccccccccc} 
\hline
Source Name              &   Class   &      L     &      B     &  $\alpha_{BLL}$ &  $\alpha_{FSRQ}$ &  $\alpha_{PSR}$&  $\alpha_{BLL}+\alpha_{FSRQ}$ & Classification \\
\hline
4FGL J0000.3-7355        &           &     307.709&     -42.730&       0.951&       0.043&       0.006&       0.994&     BLL \\
4FGL J0001.2+4741        & bcu       &     114.250&     -14.338&       0.865&       0.128&       0.007&       0.993&     BLL \\
4FGL J0001.2-0747        & bll       &      89.033&     -67.305&       0.956&       0.037&       0.007&       0.993&     BLL \\
4FGL J0001.5+2113        & fsrq      &     107.649&     -40.168&       0.016&       0.981&       0.003&       0.997&    FSRQ \\
 4FGL J0001.6-4156  & bcu     &    334.226&  -72.029&   0.965&     0.029& 0.005&  0.994&     BLL   \\
\hline
	\end{tabular}
\end{table*}

\begin{table}
	\centering
	\caption{Classification results for DR1 (8y) and DR2 (10y) sources. The first column represents the source class as assigned in the 4FGL catalogues, and the second and third columns show the number of sources for a given class in the DR1 and DR2 respectively. The following columns show the percentage of sources classified by our ML model-ensemble as BLL (column 4 DR1, column 5 DR2), FSRQ (column 6 DR1, column 7 DR2) or PSR  (column 8 DR1, column 9 DR2). Source name acronyms are listed in Appendix~\ref{app:acronyms}.}
	\label{tab:all_classes_table}
	\begin{tabular}{lrcccrccc} 
		\hline
		Class  & \multicolumn{2}{c}{Srcs} &  \multicolumn{2}{c}{BLL (\%)} & \multicolumn{2}{c}{FSRQ  (\%)}& \multicolumn{2}{c}{PSR  (\%)}   \\
		\hline	
           & 8y & 10y  & 8y & 10y  & 8y & 10y  & 8y & 10y  \\
\hline 
BLL     &    1131     &       1190     &      92    &         94      &       7       &      5         &       1       &      1           \\
FSRQ    &     694     &        730      &      13    &          15     &      86      &     83        &       1       &      1           \\
PSR     &     239     &         259     &      4.5   &           5      &      4.5     &       3        &     91        &     92          \\
BCU      &    1310     &       1515    &       61    &         62     &      36      &      31        &      3         &       7         \\
UID     &    1336     &       1679    &       30    &         37     &      38      &      21        &    32         &     42         \\
 GAL    &    3     &    5      &     33    &  60       &     0       &      0    &     67      &   40        \\ 
 CSS    &    5     &    5      &      0    &  20       &     100     &     80    &     0       &   0 	  \\ 
 AGN    &   11     &   11      &     64    &  45       &     36      &     36    &     0       &   18 	  \\ 
 SSRQ   &    2     &    2      &      0    &   0          &     100     &    100    &     0       &   0	  \\ 
 NLSY1  &    9     &    9      &     0     &   0       &     100    &    100    &     0       &    0 	  \\ 
 SEY    &    1     &    1      &     100   &  100      &     0       &      0    &     0       &    0	  \\ 
 RDG    &   42     &   44      &     71    &  77       &     24      &     18    &     5       &    5 	  \\ 
 GLC    &   30     &   30      &      3    &  10       &     20      &     10    &     77      &   80 	  \\ 
 HMB    &    8     &    8      &     37    &  25       &     25      &     25    &     37      &   50 	  \\ 
 LMB    &    2     &    3      &     0     &   67      &     50       &    33      &    50      &     0	  \\ 
 NOV    &    1     &    1      &     0     &   0       &     100      &    100    &     0       &    0 	  \\ 
 SBG    &    7     &    8      &     86    &  100      &     14      &      0    &     0       &    0	  \\ 
 SFR    &    3     &    5      &     100   &  80       &     0       &      0    &     0       &   20 	  \\ 
 SPP    &   78     &   95      &     24    &  26       &     27      &     14    &     49      &   60 	  \\ 
 SNR    &   40     &   43      &     57    &  56       &     5       &      2    &     37      &   42 	  \\ 
 PWN    &   18     &   18      &     83    &  78       &     6       &      6    &     6       &   11 	  \\ 
 UNK    &   92     &  115      &     35    &  43       &     29      &     10    &     36      &   47	  \\ 
		\hline
	\end{tabular}
\end{table}

\begin{figure}
	\includegraphics[width=\columnwidth]{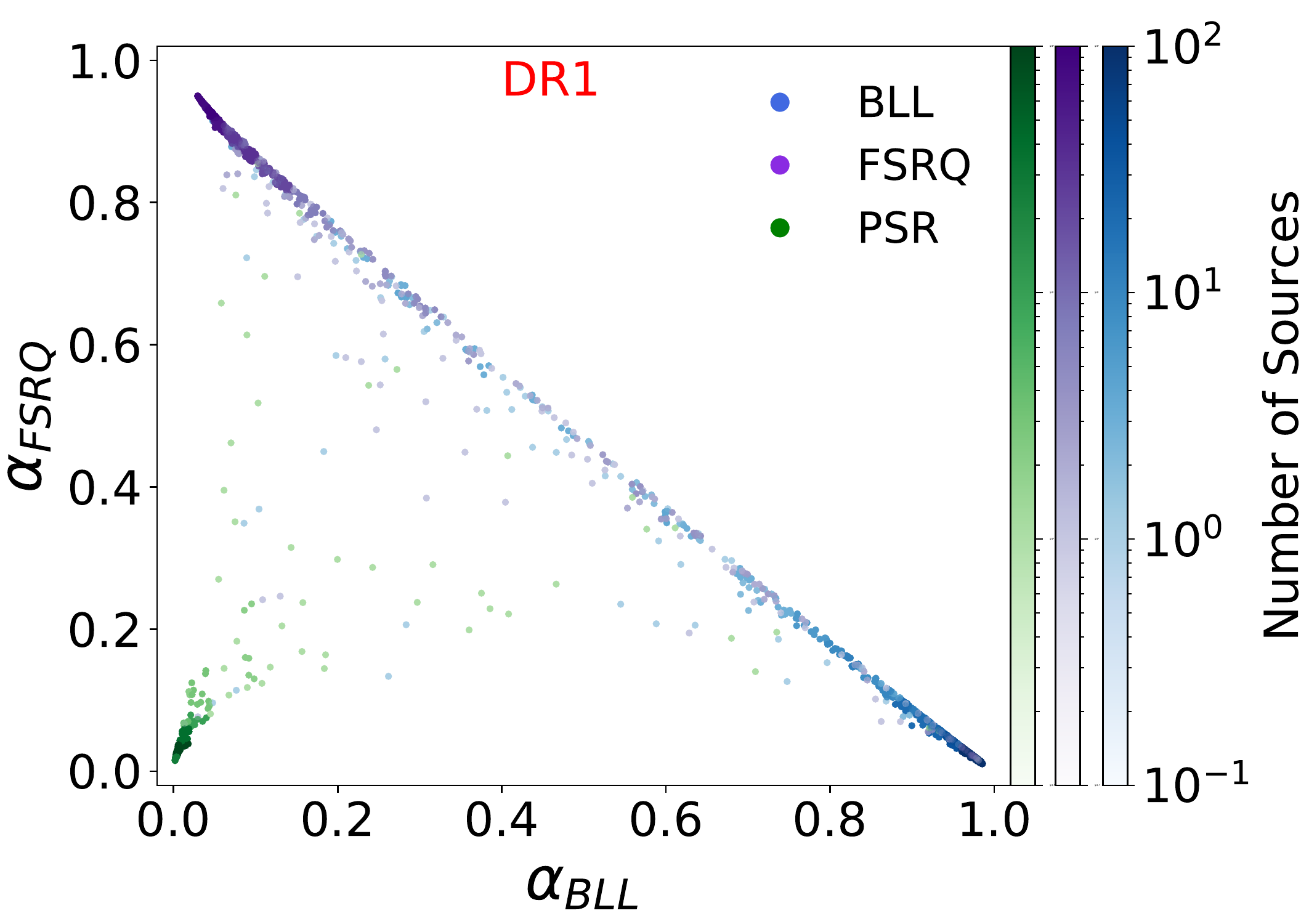}
	\includegraphics[width=\columnwidth]{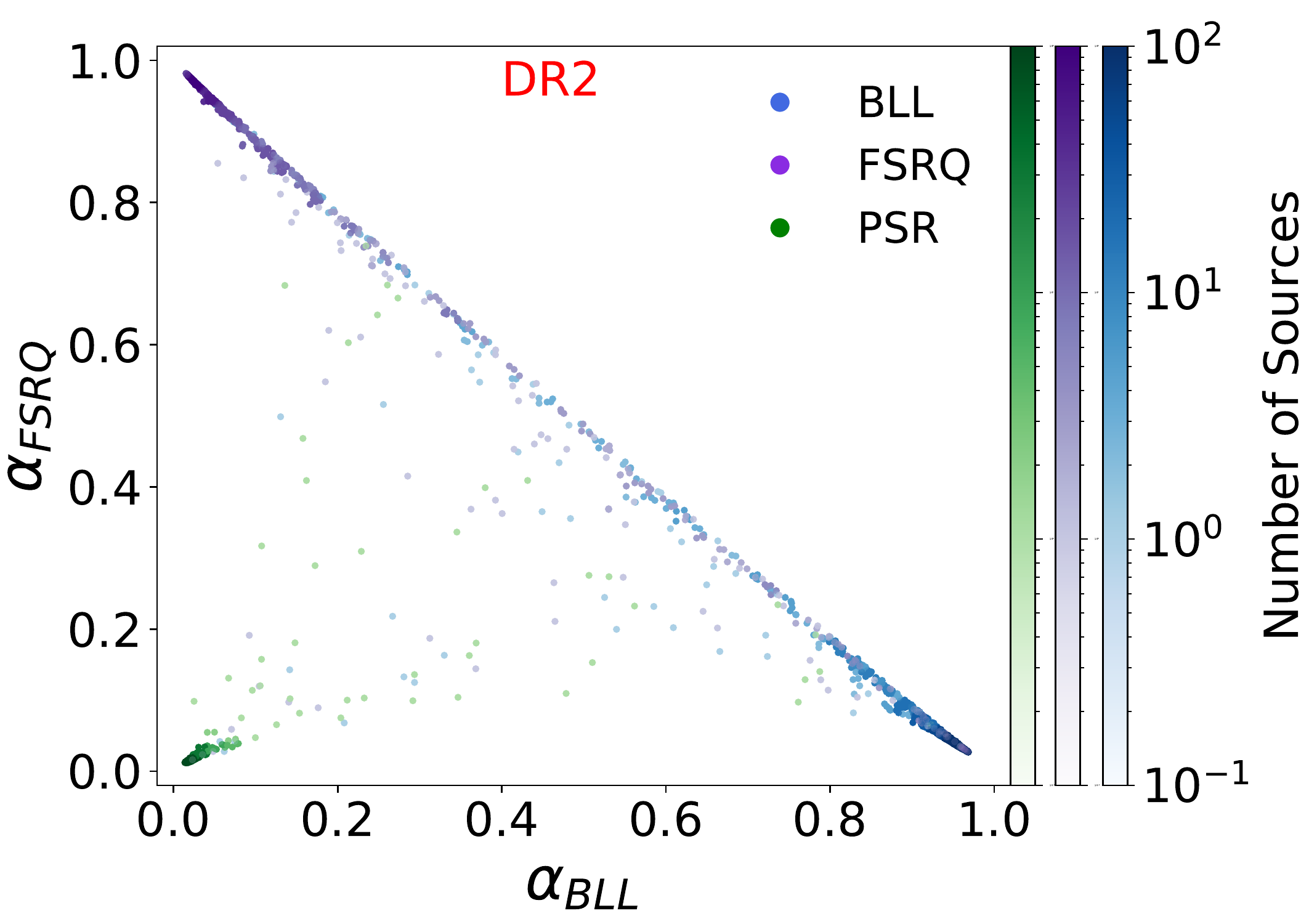}
    \caption{DR1 (top) and DR2 (bottom) distribution on the $\alpha_{BLL}-\alpha_{FSRQ}$ plane for sources belonging to the BLL (blue dots), FSRQ (purple dots) and PSR (green dots) classes. The colours represent the number of sources in the vicinity of the point. A point with very low values of $\alpha_{BLL}$ and $\alpha_{FSRQ}$  corresponds to a source with a high likelihood of being a Pulsar ($\alpha_{BLL}+\alpha_{FSRQ}+\alpha_{PSR} = 1$).}
  \label{fig:classification_main_cl}
\end{figure}

Table~\ref{tab:all_classes_table} shows a summary of the classification results in terms of source classes. 
The first thing to note is that a fraction (77\%) of the sources belonging to the first three rows are used for the ANN training, results about the validation sample are discussed in Section \ref{sec:validation}.
The classification likelihoods for sources belonging to those three classes are shown on a $\alpha_{BLL}-\alpha_{FSRQ}$ plane in Figure~\ref{fig:classification_main_cl}.
Since the sum of the three $\alpha$ parameters is equal to one, sources with a high value of $\alpha_{PSR}$ are clustered close to the axes origin.  Table~\ref{tab:all_classes_table} and Figure~\ref{fig:classification_main_cl} both show comparable results for DR1 and DR2. The plots' main features are the large separation of the bulk of the Pulsars from  BLL or FSRQ and the diagonal band showing that some Blazars are firmly classified as non-PSR, but the likelihood of belonging to one of the two Blazar subclasses is not sharply defined.

\subsection{Validation}
\label{sec:validation}


The ANN classification results for the validation sample are shown in Table~\ref{tab:test_table}, the three kinds of sources are on average correctly classified with comparable performance to that shown in Table~\ref{tab:all_classes_table}.

\begin{table}
	\centering
	\caption{Classification results for DR1 (8y) and DR2 (10y) sources belonging to the BLL, FSRQ and PSR classes not used for training (validation sample). The first column represents the source class as assigned in the 4FGL catalogues, and the second and third columns show the number of sources for a given class in the DR1 and DR2 respectively. The following columns show the percentage of sources classified by our ML model-ensemble  as BLL (column 4 DR1, column 5 DR2), FSRQ (column 6 DR1, column 7 DR2) or PSR (column 8 DR1, column 9 DR2).}
	\label{tab:test_table}
	\begin{tabular}{lrcccrccc} 
		\hline
		Class  & \multicolumn{2}{c}{Srcs} &  \multicolumn{2}{c}{BLL (\%)} & \multicolumn{2}{c}{FSRQ  (\%)}& \multicolumn{2}{c}{PSR  (\%)}   \\
		\hline	
           & 8y & 10y  & 8y & 10y  & 8y & 10y  & 8y & 10y  \\
\hline 
BLL    &  349  &  391  &  89.4 & 95.5   &   10.3   &   4   &   0.3  &  0.5   \\
FSRQ   &  249  & 252  &  13.6 &  17      &   86     &  81   &   0.4  &   2    \\ 
PSR    &  84    &   77  &     5   &    8      &    7      &   1    &   88   &  91    \\
\hline
	\end{tabular}
\end{table}

\begin{figure}
	\includegraphics[width=\columnwidth]{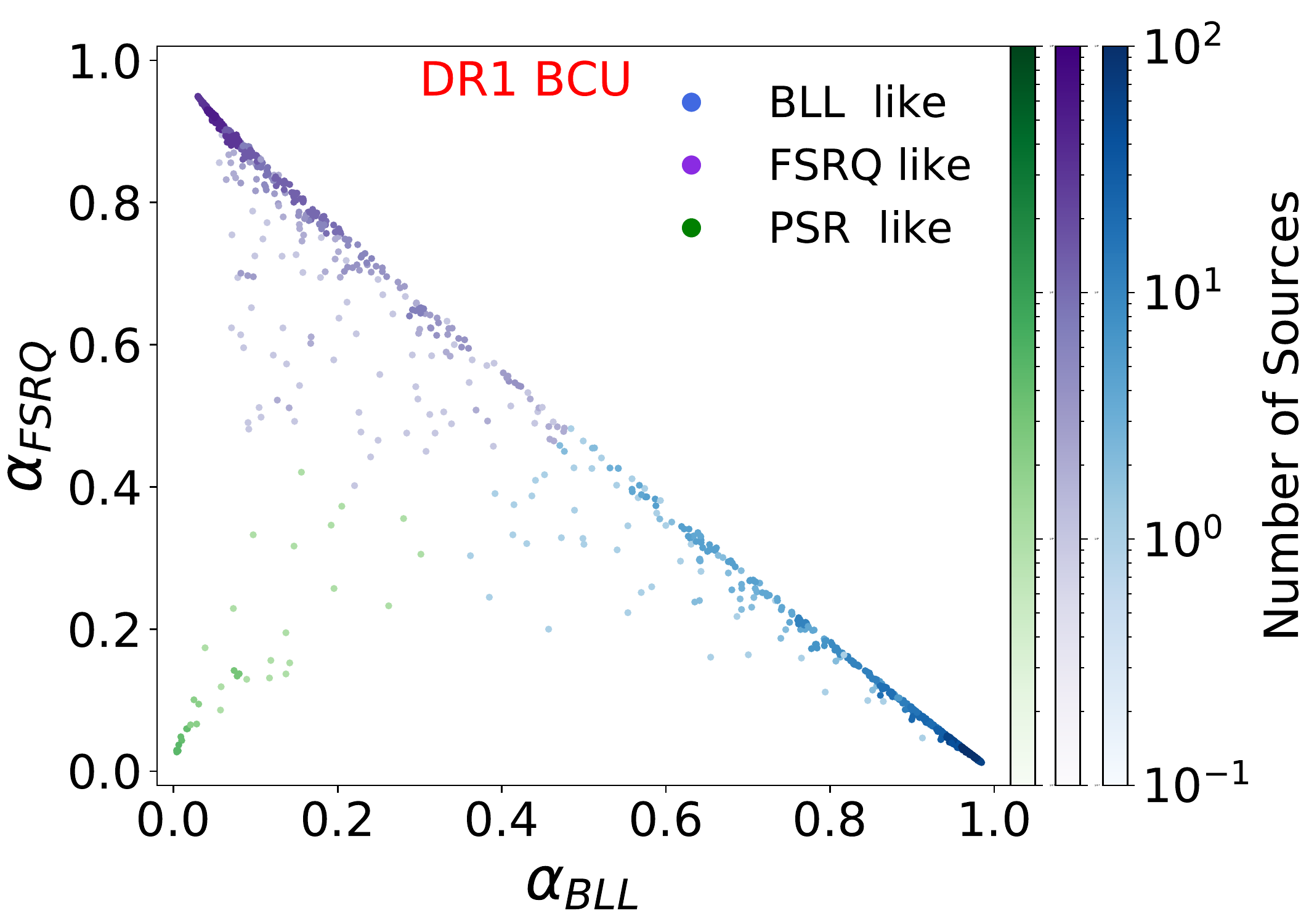}e
	\includegraphics[width=\columnwidth]{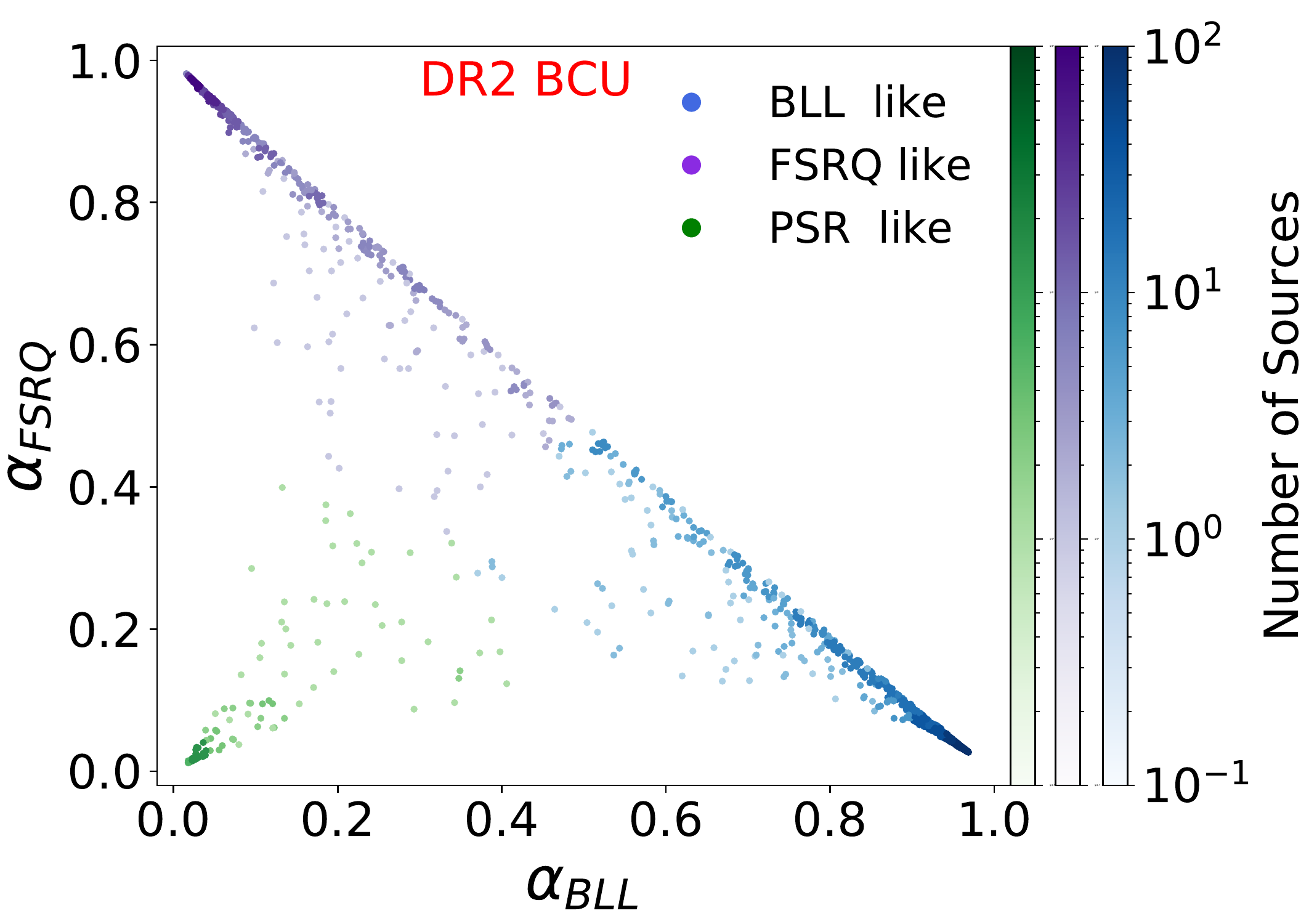}
    \caption{DR1 (top) and DR2 (bottom) distribution on the $\alpha_{BLL}-\alpha_{FSRQ}$ plane BCU sources.  The colours represent the number of sources in the vicinity of the point. Blue, purple and green dots correspond to sources classified respectively as most probably BLL, FSRQ and PSR.}
  \label{fig:classification_bcu}
\end{figure}

Although BCU sources are not used within training, Table~\ref{tab:all_classes_table} shows that they are mainly classified as either BLL or FSRQ; therefore, they are correctly identified as blazars in more than 90\% of cases for both DR1 and DR2.  
Additionally, the fraction of BCU classified as BLL or FSRQ is consistent between DR1 and DR2.
Figure~\ref{fig:classification_bcu} shows the BCU classification results on the $\alpha_{BLL}-\alpha_{FSRQ}$ plane, as for BLL and FSRQ, they are mainly distributed on a diagonal band characterising them as blazars even when the classification algorithm cannot distinguish between BLL and FSRQ. Together with the three   $\alpha$ parameters already introduced, we can define the blazar likelihood parameter $\alpha_{BLZ} = \alpha_{BLL} + \alpha_{FSRQ}$ to identify all the sources lying on the blazar diagonal band independently from their class type (BLL or FSRQ).

\begin{figure}
	\includegraphics[width=\columnwidth]{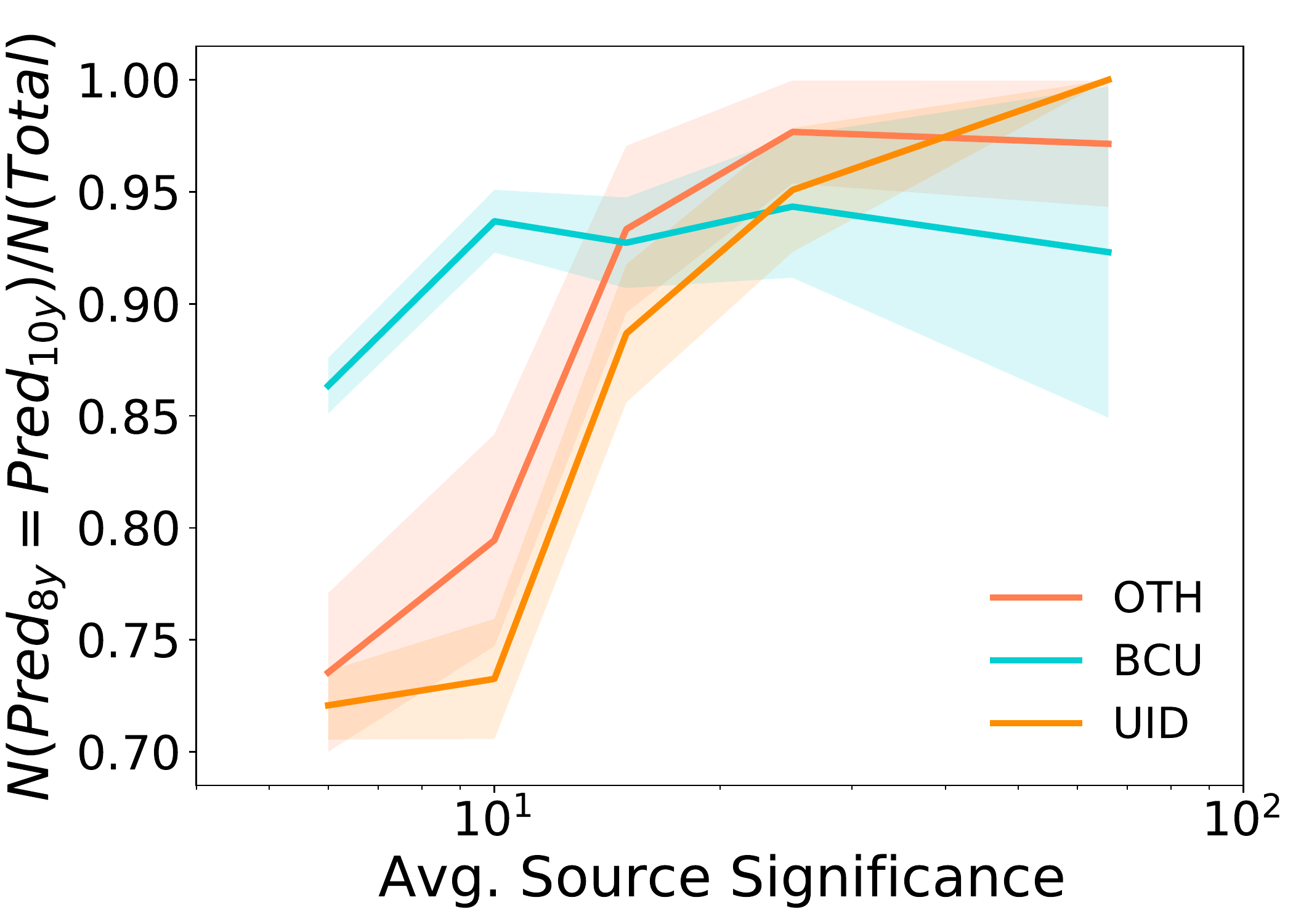}
    \caption{DR1 and DR2 agreement fraction on the most probable classification for UID, BCU and OTH sources as a function of the source significance.  The band represents the binomial error, while the source significance is taken as the average between the DR1 and DR2 values.}
  \label{fig:8y_10y_agreement_fraction}
\end{figure}

All the sources in the 4FGL DR1 are also present in the DR2 even if 120 of them are formally below the detection threshold, Figure~\ref{fig:8y_10y_agreement_fraction} shows the fraction of sources for which DR1 and DR2 ANNs agree on the most probable category considering all the sources belonging to classes not used for training; BCU, UID and OTH sources are shown separately. In all cases the agreement fraction improves with the source significance (column \textit{Signif\_Avg } in the 4FGL catalogue), with values always greater than 70\% (85\% for BCU) and reaching above 90\% at high significance levels. 

\begin{figure}
	\includegraphics[width=\columnwidth]{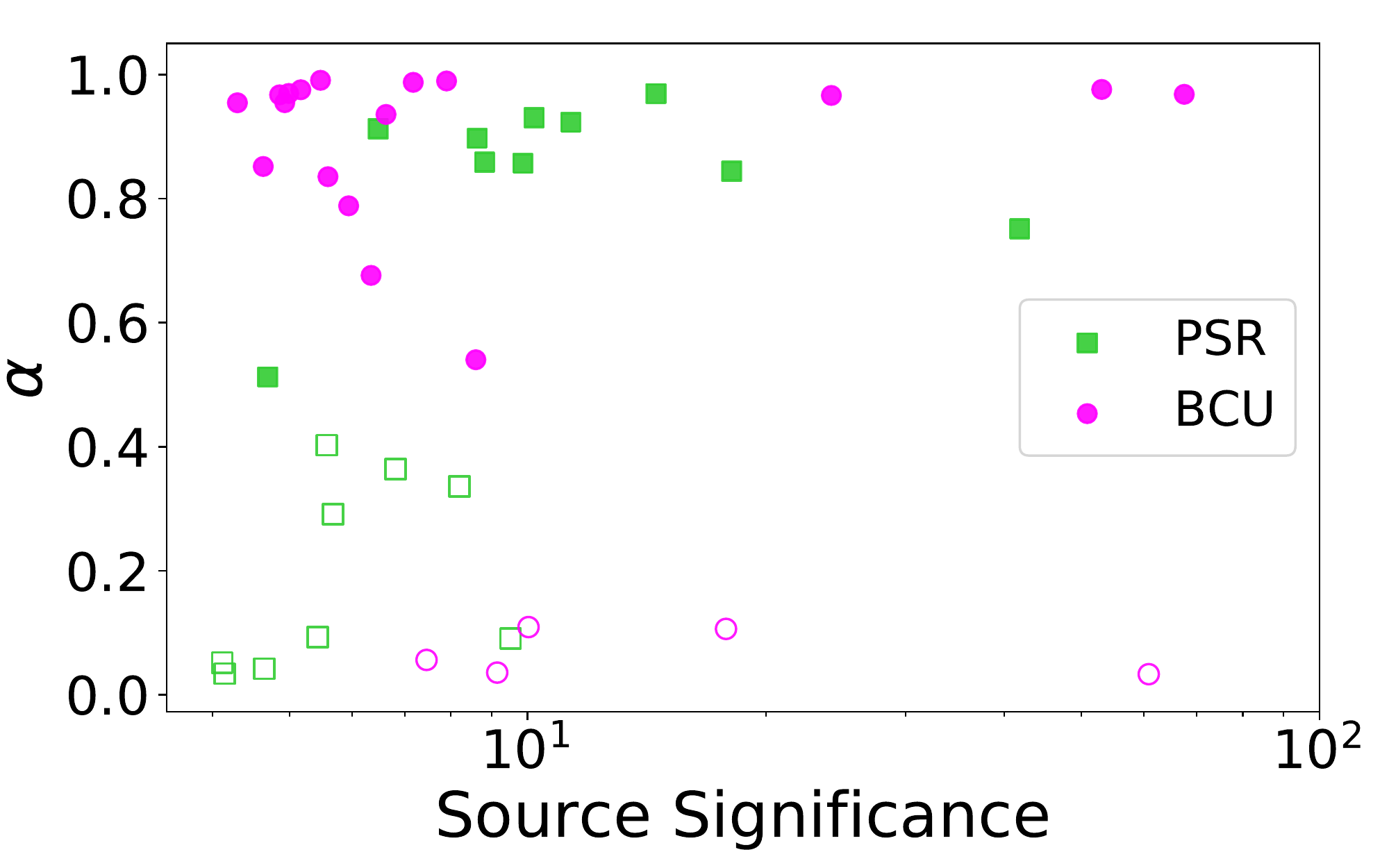}
    \caption{Correct association likelihood ($\alpha$) as a function of the significance for sources whose Class changed into PSR in DR2 (green squares) or into BCU (magenta circles). The association likelihood is equal to $\alpha_{PSR}$ and $\alpha_{BLL}+\alpha_{FSRQ}$ for Pulsars and BCU, respectively. Full (empty) symbols are sources whose DR1 classification matches (does not match) DR2 Class.}
  \label{fig:8y_10y_class_change}
\end{figure}

Among the common sources, 58 are associated with a different class in DR1 and DR2, 40 are unassociated in DR1. None of the sources moved into the BLL or the FSRQ class. 
Figure~\ref{fig:8y_10y_class_change} shows the relevant $\alpha$ classification parameter for the sources whose class changed into PSR (19) or BCU (22) in the DR2 catalogue. DR1 classification tends to match the new DR2 class for higher significance sources with two exceptions of note (4FGL J1745.6-2859, significance 67.5 and 4FGL J1102.0-6054, significance 17.8) which are both labelled as sources with a potential association with a SNR or a PWN (SPP) in the DR1 and to a BCU in DR2. They are classified as PSR-like in both the DR1 and DR2. SPP are classified as PSR-like in 49 \% (60\%) of the cases in DR1 (DR2), while less than 10\% of the BCU are misclassified as pulsars. This hints at the possibility of the DR1 association being more valid than the DR2 association for those two sources.

\subsection{Unassociated Sources}

\begin{figure}
	\includegraphics[width=\columnwidth]{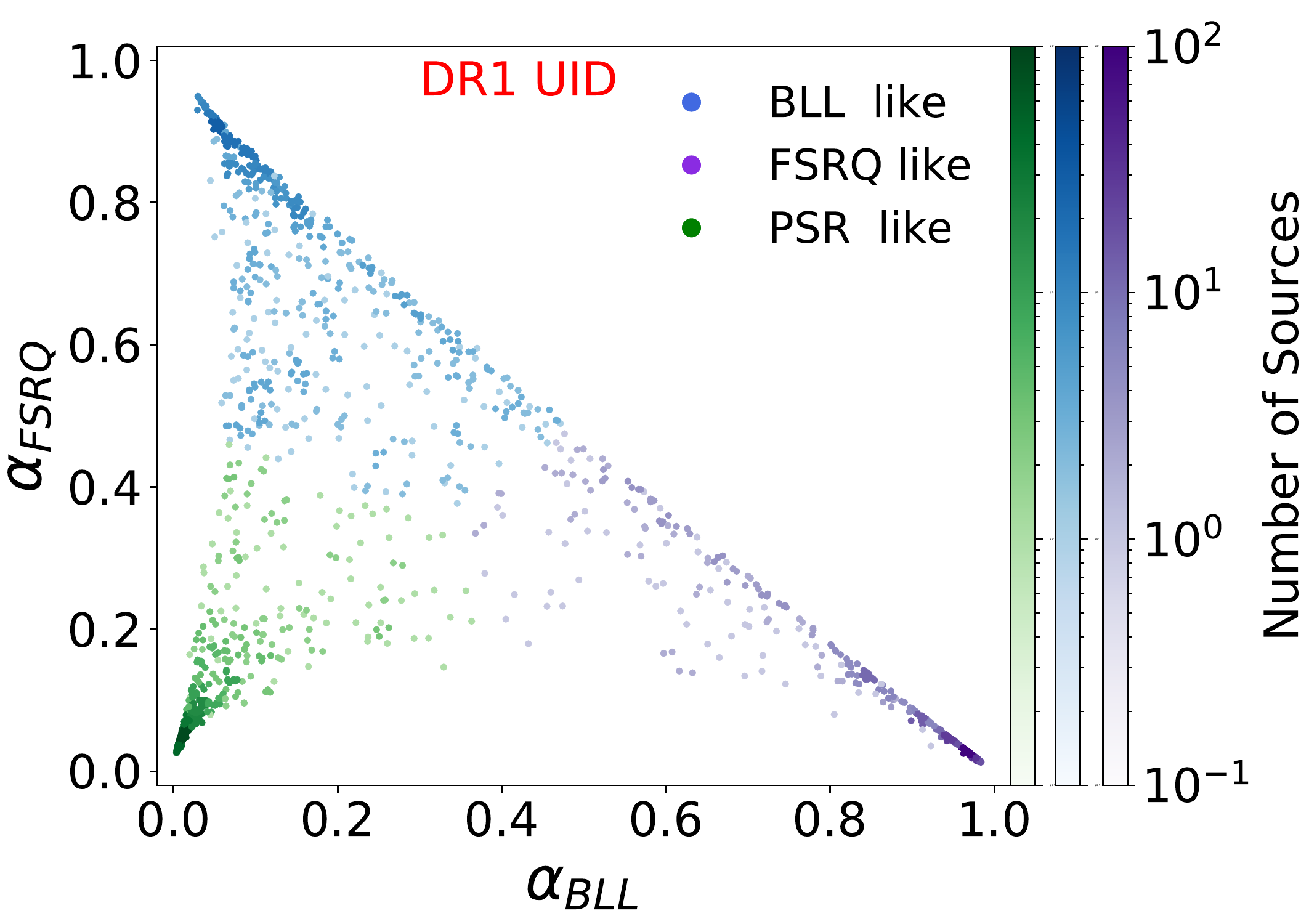}
	\includegraphics[width=\columnwidth]{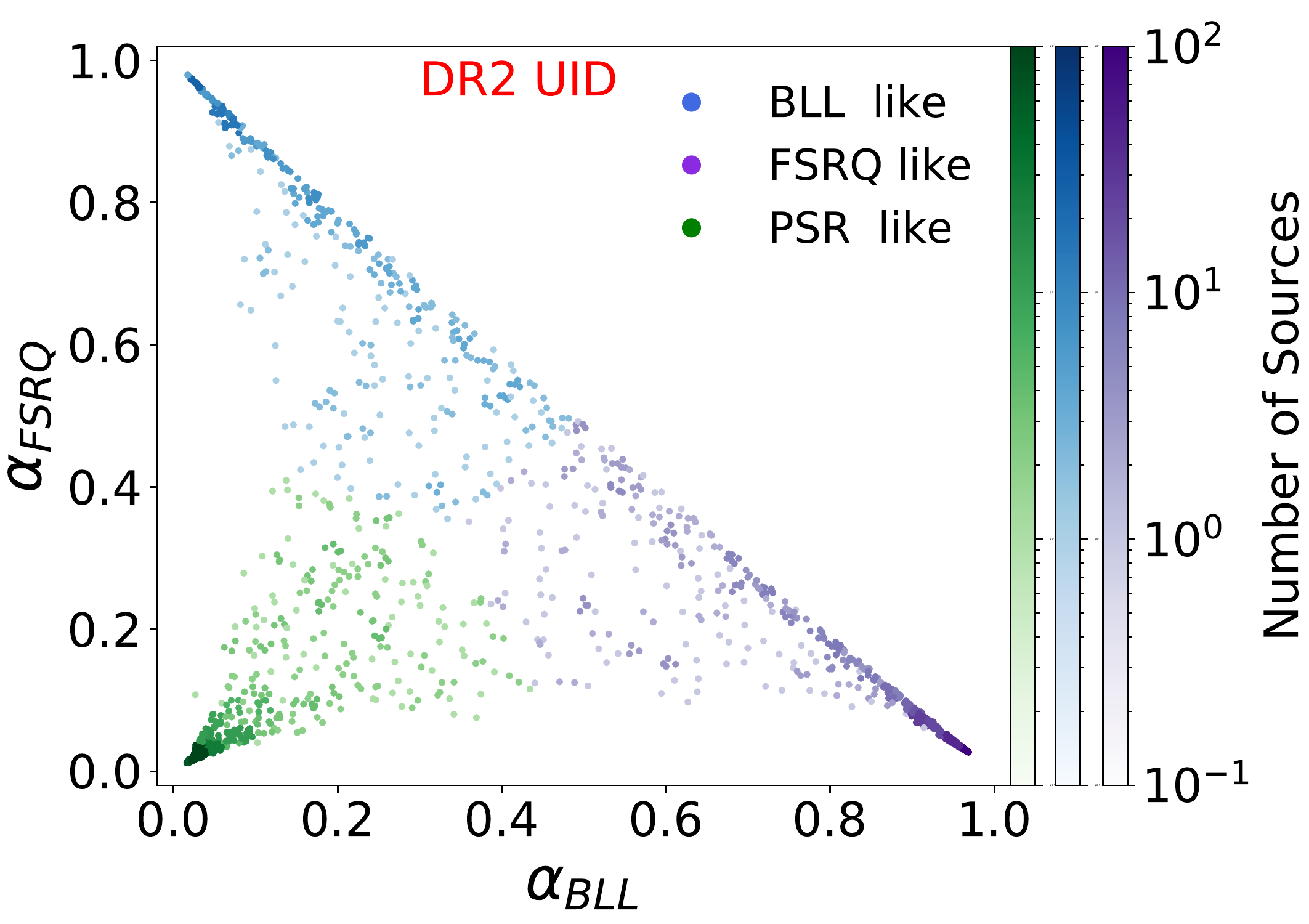}
    \caption{DR1 (top) and DR2 (bottom) distribution on the $\alpha_{BLL}-\alpha_{FSRQ}$ plane for unassociated sources. The colours represent the number of sources in the vicinity of the point. Blue, purple and green dots correspond to sources classified respectively as most probably BLL, FSRQ and PSR.}
  \label{fig:classification_uid}
\end{figure}

The ANN classification of 4FGL unassociated sources  (see also Table~\ref{tab:all_classes_table}) resulted in rough equipartition among BLL, FSRQ and PSR with DR2 classification predicting a slightly higher number of pulsars and BLL with a reduced number of FSRQ. The equiprobability of ANN classification is in sharp contrast with the population's asymmetric size for the corresponding identified sources (see also Table~\ref{tab:4FGL}).
The UID distribution on the $\alpha_{BLL}-\alpha_{FSRQ}$ plane  is shown in Figure~\ref{fig:classification_uid}. As for BLL, FSRQ, PSR (Figure~\ref{fig:classification_main_cl}) and BCU (Figure~\ref{fig:classification_bcu}), most of the sources cluster at the vertices of the triangular distribution where the three $\alpha$ parameters are close to one. The distinctive blazar diagonal band is present too, but a  higher fraction of sources lie in the central region with no strong association to a single category.

\begin{figure}
	\includegraphics[width=\columnwidth]{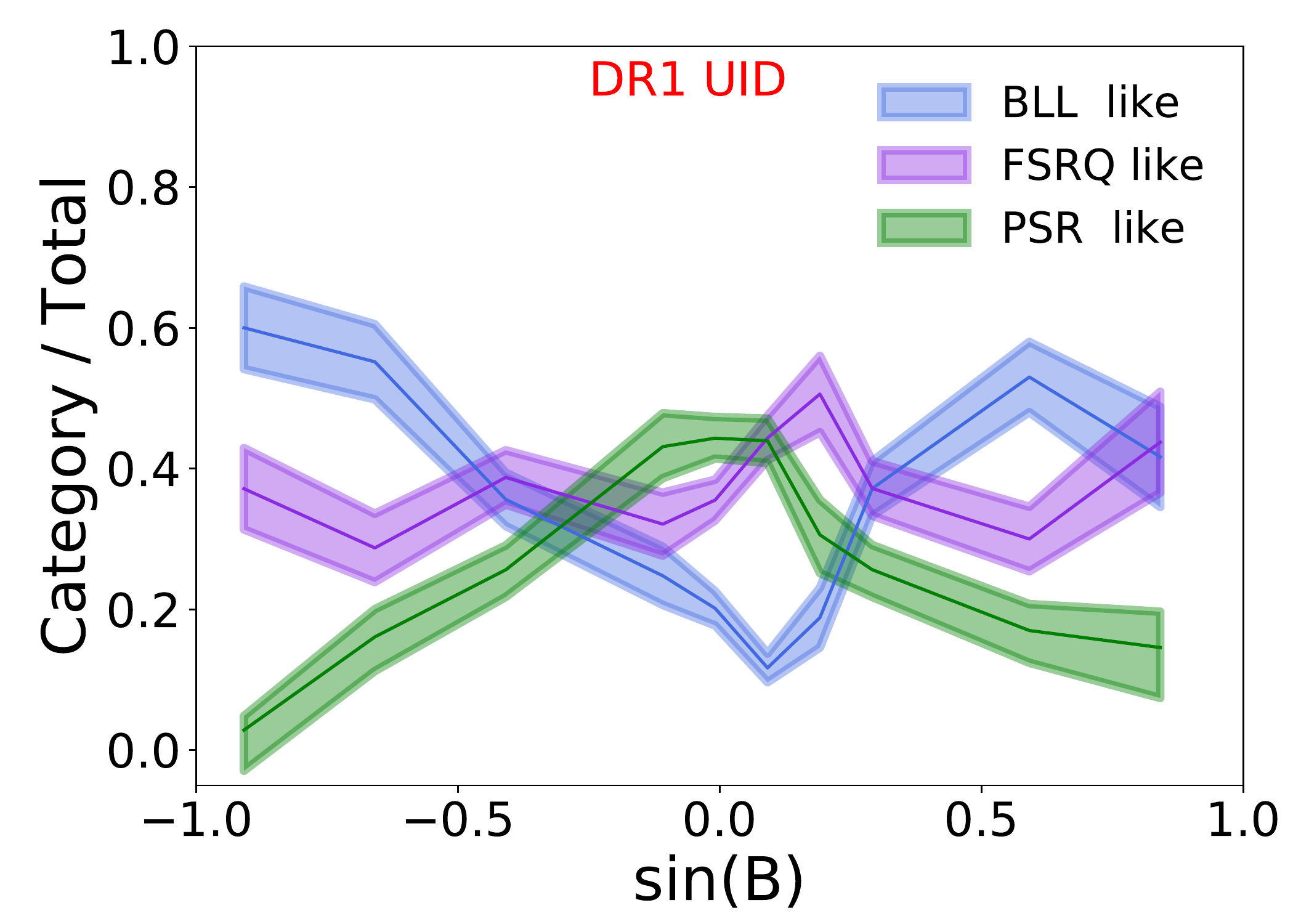}
	\includegraphics[width=\columnwidth]{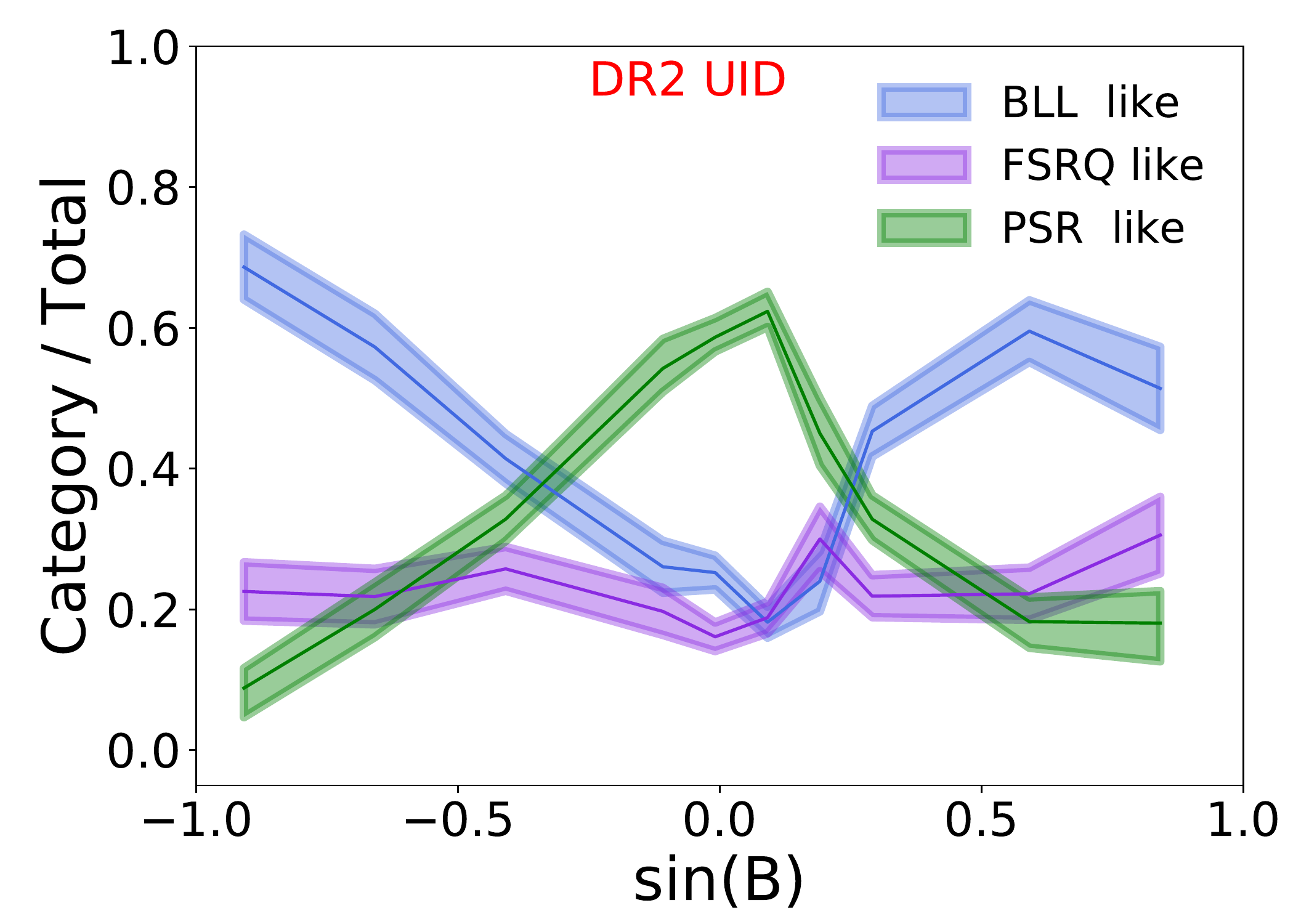}
   \caption{DR1 (top) and DR2 (bottom) fraction of sources as a function of the sinus of the galactic latitude $B$ for the three classification categories.}
  \label{fig:classification_uid_sinb}
\end{figure}

Galactic latitude ($B$) is not part of the information used for the ANN training. However, by analysing the UID classification results as a function of $sin(B)$ in Figure~\ref{fig:classification_uid_sinb} we see that PSR-like sources are dominant at low latitude and the relative fraction tends to zero at high latitude for both DR1 and DR2. This is expected for a correct classification since Pulsars are galactic sources which, as shown by~\citet{4FGL}  and~\citet{FermiL2PSR}, are mainly present at a low galactic latitude with a sub-population of Millisecond Pulsars (MSP) spreading from the galactic plane. 

Extragalactic sources are expected to be uniformly distributed across the sky even if, as shown by~\citet{4LAC},  the galactic latitude distribution of associated blazars shows a reduced number   close to the galactic plane because of 
 the higher level of diffuse background which masks the fainter extragalactic sources.
 We  found (Figure~\ref{fig:classification_uid_sinb}) that the FSRQ-like fraction of UID is roughly constant across all values of $B$, while BLL-like sources dominate at high galactic latitudes, both this behaviours can be considered reasonable.
The origin of  this difference between the UID classification into the two blazar subclasses is unclear but  it is worth mentioning that, as discussed in Section~\ref{sec:oth}, unidentified sources belonging to the OTH classes are being forced into PSR-like, BLL-like or FSRQ-like.
A second order effect on the galactic latitude distribution of the associated Fermi blazars, also discussed in ~\citet{4LAC},   is related to the catalogues used for the association with a resulting asymmetry between the number of associated BLL at North and South galactic latitudes and an excess of associated BLL for $sin(B)$ approaching the value of $+1$, which  could explain the corresponding dip in the UID BLL-like fraction present in both the DR1 and DR2 classification results.

The fact that the  galactic latitude distributions of the UID classification conform to the expectations is  a reassuring result of the classification's general behaviour and its predictive potential for unassociated sources.

\subsection{Other Sources}
\label{sec:oth}

\begin{figure}
	\includegraphics[width=\columnwidth]{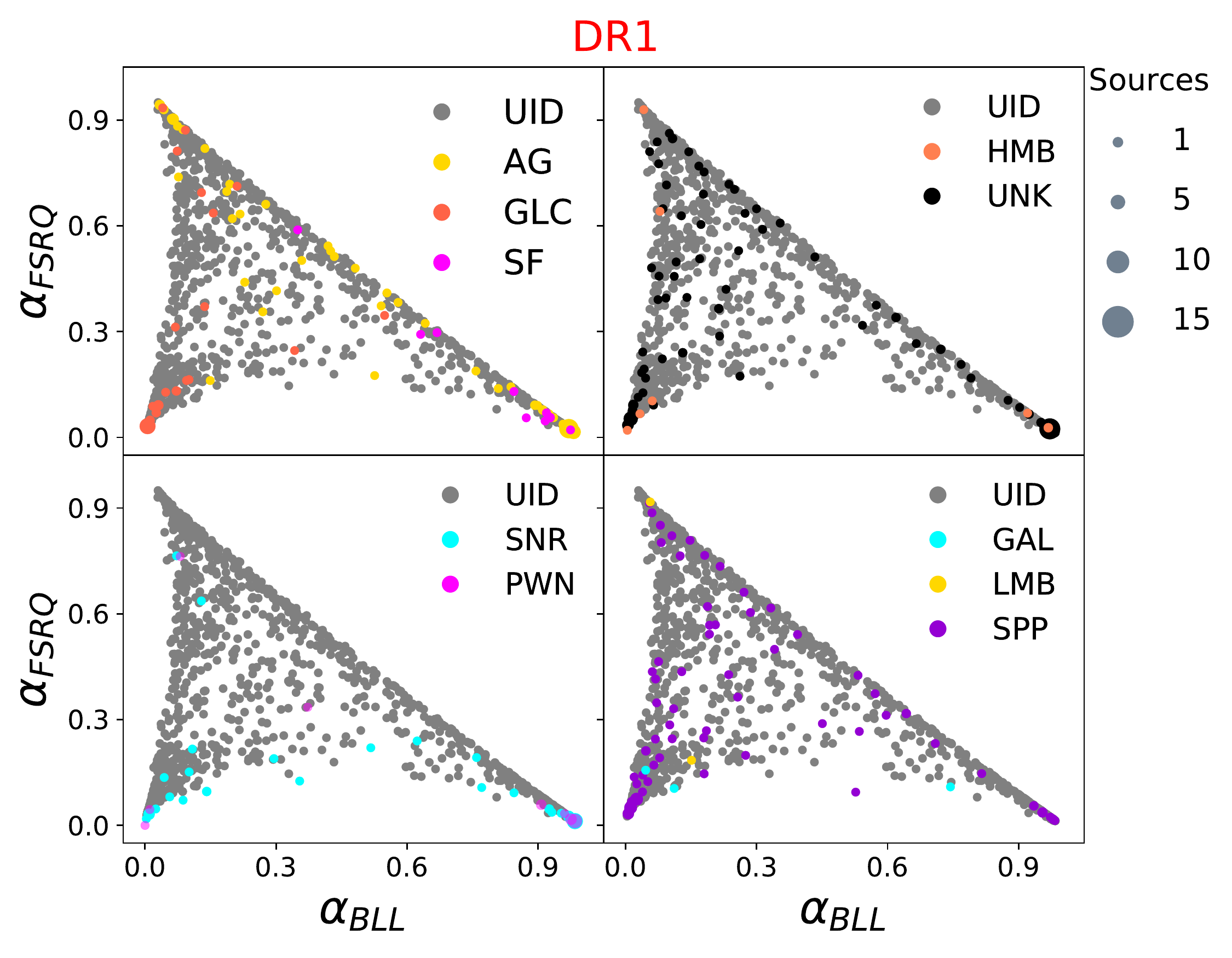}
	\includegraphics[width=\columnwidth]{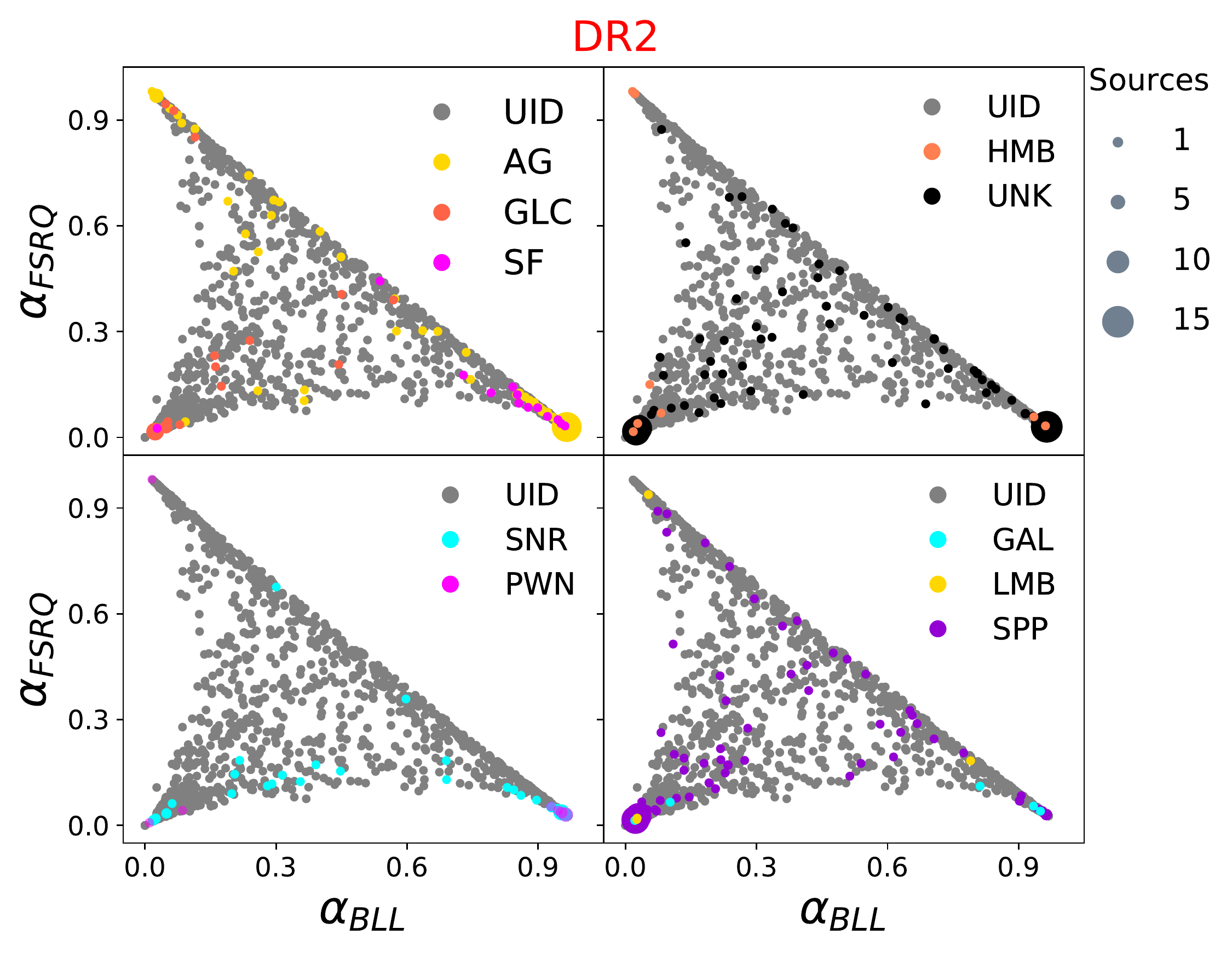}
   \caption{DR1 (top) and DR2 (bottom) distribution on the $\alpha_{BLL}-\alpha_{FSRQ}$ plane. Grey dots represent Unassociated sources in all quadrants. The dot size for associated sources (non-grey) corresponds to the number of sources in the vicinity. Yellow dots in the top left quadrants show all non-Blazar sources related to active galaxies AGN, CSS, RDG, SSRQ, NLSY1, SEY). In the same quadrant, orange and magenta dots represent Globular Clusters and sources related to high star formation rate (SFR, SBG), respectively. In the bottom left quadrants, cyan and magenta dots correspond to Supernova Remnants and Pulsar Wind Nebulae, respectively. The bottom right quadrants show Galaxies (cyan dots), Low Mass Binaries (yellow dots) and Sources potentially associated with SNR or PWN (violet dots). In the top-right quadrants, orange dots represent High Mass Binaries, and black dots correspond to sources of unknown type.}
  \label{fig:classification_oth}
 \end{figure}


Analysing the classification results for all the source classes through Table~\ref{tab:all_classes_table}  and Figure~\ref{fig:classification_oth} it is also possible to search for patterns present in the UID classification. Different source classes cluster in different areas of the $\alpha_{BLL}-\alpha_{FSRQ}$ plane with a common classification behaviour between DR1 and DR2. This indicates that, even if the classification cannot point uniquely to one of the OTH subclasses, the source position in the plane is related to its gamma-ray properties. 

Sources related to active galaxies other than blazars (upper left quadrants) are typically classified as BLL and are almost exclusively distributed on the blazar diagonal band. 
Gamma-ray emission from Globular Clusters (GLC) is expected to come from millisecond pulsars~\citep[][]{Fermi_GLCpop, Fermi_GLCpsr} which are mainly classified as PSR-like sources; indeed this is an indication that the ANN results are robust. By grouping together star formation regions and starburst galaxies it appears that gamma-ray sources related to high star formation rates tend to be classified as BLL which can be related to their hard, non-variable, power-low spectrum~\citep[][]{Fermi_SFgal}.

Pulsar Wind Nebulae (PWN), in the bottom right quadrants, have mostly a high value of  $\alpha_{BLL}$, while Supernova Remnants (SNR)  form an arch between the two points with $\alpha_{PSR}=1$ and  $\alpha_{BLL}=1$, corresponding to the common feature of a low value for $\alpha_{FSRQ}$. This can be explained with a non-variable power-law spectrum where in some cases a   \emph{LogParabola} curvature is observable~\cite[][]{Fermi_SNRCat1}; the SNR with significant curvature tend to be classified as PSR-like. 

Sources with a potential association with SNR or PWN (SPP) in the bottom right quadrants, partially mimic the distribution of their potential associated classes, but with a disproportionate fraction featuring a high  $\alpha_{PSR}$, as evident from Table~\ref{tab:all_classes_table}, where SPP show a higher PSR-like fraction than either SNR or PWN. Some SPP are also classified with a relatively large value of $\alpha_{FSRQ}$ in contrast with the main populations of SNR and PWN.  
Contrary to the other gamma-ray known classes not related to  AGN, SPPs show several sources lying on the blazar diagonal band. These features could be a hint that some of the potential associations among SPP are incorrect. 
A direct comparison of SPP with SNR and PWN in terms of spectral or time variability properties show that the main difference is in the power-law spectral index (\textit{PL\_index}); almost all PWN and SNR feature a hard spectrum with a spectral index below 2.5 while a large fraction of SPP, including most of those classified with high $\alpha_{FSRQ}$, show a value of \textit{PL\_index} greater than 2.5.

Galaxies are classified either as PSR-like or BLL-like, with the few Low-Mass Binaries (LMB) not belonging consistently to any singular category.

Sources of unknown type in the top right quadrants do not show a specific behaviour, as expected since they probably belong to more than one source class. High-Mass Binaries (HMB) are not preferentially classified in a specific category.
Interestingly, the two OTH subclasses whose astrophysical characterisation is not certain (SPP and UNK) show a non-negligible fraction of sources that populate the plot's middle ground where none of the $\alpha$ parameters is high or low. In contrast, the other sources' classification usually results in at least one of the $\alpha$ values being close to zero.

\subsection{Source Significance}
\label{sec:alpha_vs_signif}

\begin{figure*}
	\includegraphics[width=\columnwidth]{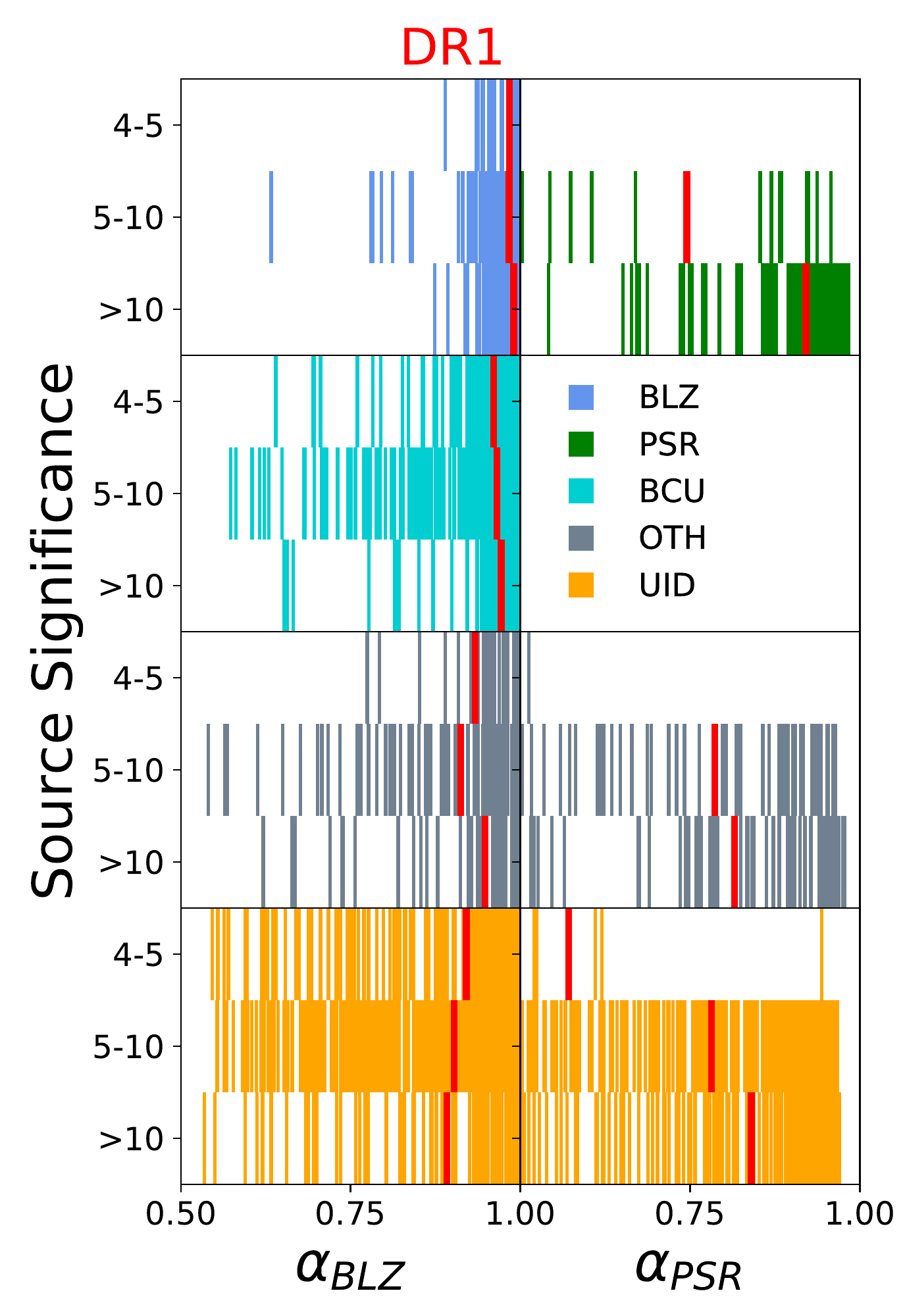}
	\includegraphics[width=\columnwidth]{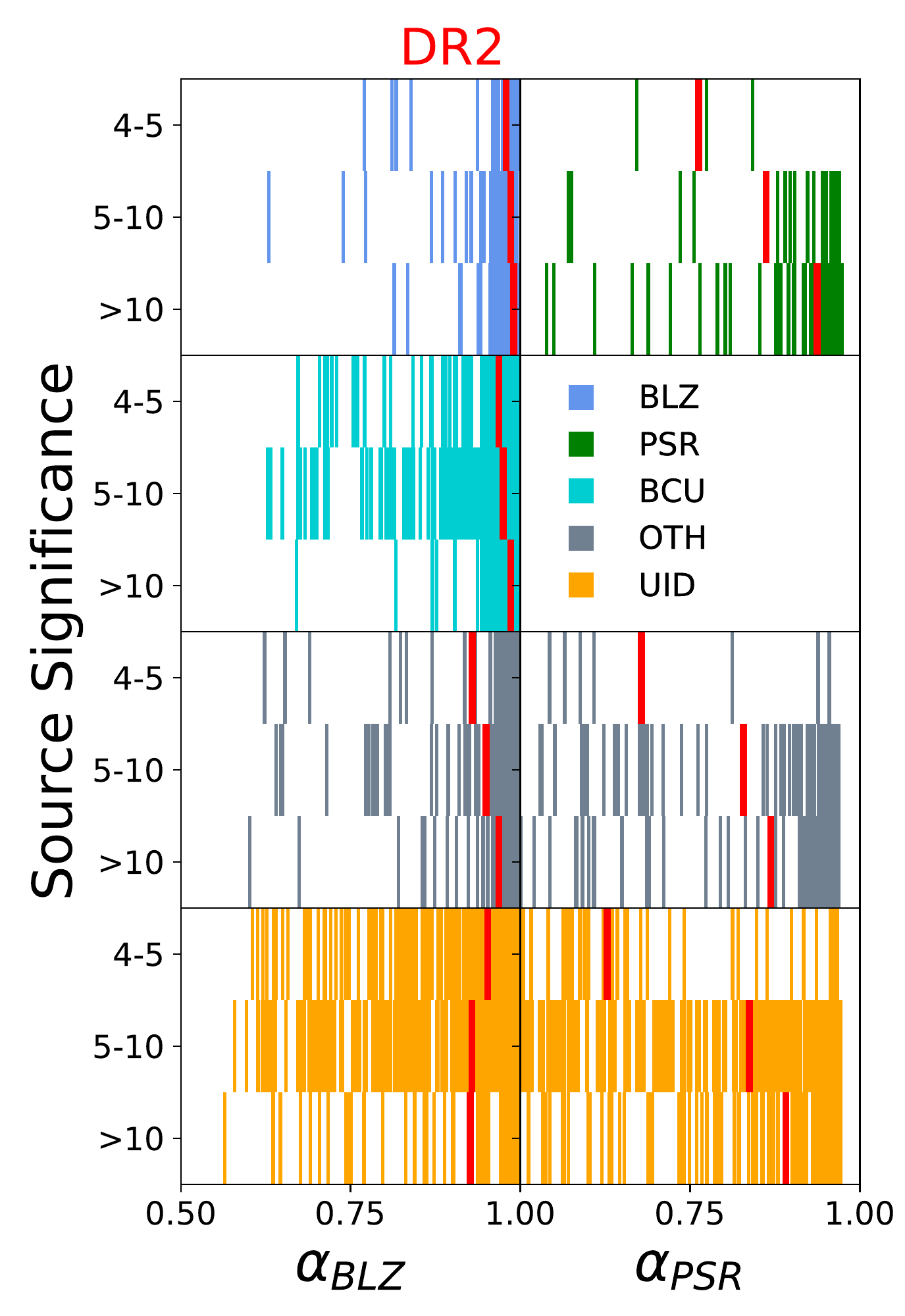}
   \caption{DR1 (left plot) and  DR2 (right plot) $\alpha_{BLZ}$ (left quadrants) and $\alpha_{PSR}$ (right quadrants) for different source Significance intervals and source classes.  $\alpha_{BLZ}$ is shown only for Blazar-like sources and $\alpha_{PSR}$ only for Pulsar like ones. The red lines show the mean of the relevant $\alpha$ parameter. The top row left and right quadrants show BLZ (blue) and PSR (green). The second and third rows from the top quadrants show BCU (cyan) and OTH (grey) sources. The bottom rows (orange) present the results of the unassociated sources.}
  \label{fig:alpha_vs_signif_range}
\end{figure*}

Low flux sources may be affected by statistical fluctuations in their spectrum or light curve evaluations, producing noisy results in the classification, as suggested by the DR1 and DR2 classification agreement increasing with the significance (see Section~\ref{sec:validation} and Figure~\ref{fig:8y_10y_agreement_fraction}).

In order to search for features or anomalies in the classification results that might appear only for brighter sources, we binned the source significance values in three principal subgroups and plotted these against the values of $\alpha_{BLZ}$ and $\alpha_{PSR}$ for each class as shown in Figure~\ref{fig:alpha_vs_signif_range}.
We observe a common pattern for DR1 and DR2. Well classified BLZ and BCU have a high $\alpha_{BLZ}$ value, with the mean (the red line) always above 0.95 and increasing with the source significance. 
As already discussed in Section~\ref{sec:approach}, sources associated with pulsars tend to have a higher than average significance; thus very few members of this class are present in the lowest significance range, but the $\alpha_{PSR}$ mean value still improves with the significance. BCUs follow a pattern similar to the BLZ class but with slightly lower values for the $\alpha_{BLZ}$ parameter.  

Sources grouped into the OTH category are relatively heterogeneous, but the distribution of $\alpha$ parameters for PSR-like and BLZ-like sources follow a pattern similar to that of PSR and BLZ or BCU, respectively. 
The $\alpha_{PSR}$ show a behaviour similar to the other classes for PSR-like UID. Contrary to the other cases, the mean $\alpha_{BLZ}$ decreases with significance for BLZ-like UID (bottom left quadrants of Figure~\ref{fig:alpha_vs_signif_range}). 
Therefore there are a peculiar fraction of BLZ-like unassociated sources with a high significance and unusual characteristics causing a shift of the $\alpha_{BLZ}$ mean toward lower values. 

\subsection{Special Cases}
\label{sec:special_cases}

We analysed the UID sources with high significance and intermediate $\alpha_{BLZ}$ values to search for potentially interesting targets for follow-up studies. We applied the following  selection to define the sample of outliers:
\begin{enumerate} 
\item  $(\alpha_{BLL} > \alpha_{PSR})$  OR  $(\alpha_{FSRQ} > \alpha_{PSR})$; 
\item  4FGL $Signif\_Avg > 10$;
\item  UID $\alpha_{BLZ}$  smaller than the mean OTH value  (DR1: 0.95, DR2: 0.97);
\item  4FGL  $Flags = 0$;
\end{enumerate}
The 10 resulting outlier sources in both DR1 and DR2 are listed in Table~\ref{tab:uid_outliers}.

\begin{table}
	\centering
	\caption{UID sources selected as outliers}
	\label{tab:uid_outliers}
	\begin{tabular}{lcccc} 
\hline
Source Name              &     L     &      B   &  DR1 $\alpha_{BLZ}$ &  DR2 $\alpha_{BLZ}$  \\
\hline
4FGL J0540.0-7552        &     287.282&     -30.582&       0.900&       0.900 \\
4FGL J1015.5-6030        &     284.866&      -3.218&       0.938&       0.743 \\
4FGL J1514.8+4448        &      74.228&      56.416&       0.942&       0.953 \\
4FGL J1517.7-4446        &     328.428&      10.733&       0.695&       0.862 \\
4FGL J1635.3+4258        &      67.641&      42.461&       0.870&       0.845 \\
4FGL J1709.4-0328        &      17.441&      20.771&       0.886&       0.751 \\
4FGL J1749.8-0303        &      23.017&      12.221&       0.620&       0.646 \\
4FGL J1804.4-0852        &      19.619&       6.237&       0.933&       0.844 \\
4FGL J1816.7+1749        &      45.339&      15.506&       0.931&       0.564 \\
4FGL J1953.5+3841        &      73.837&       5.686&       0.876&       0.831 \\
\hline
	\end{tabular}
\end{table}

\begin{figure}
	\includegraphics[width=\columnwidth]{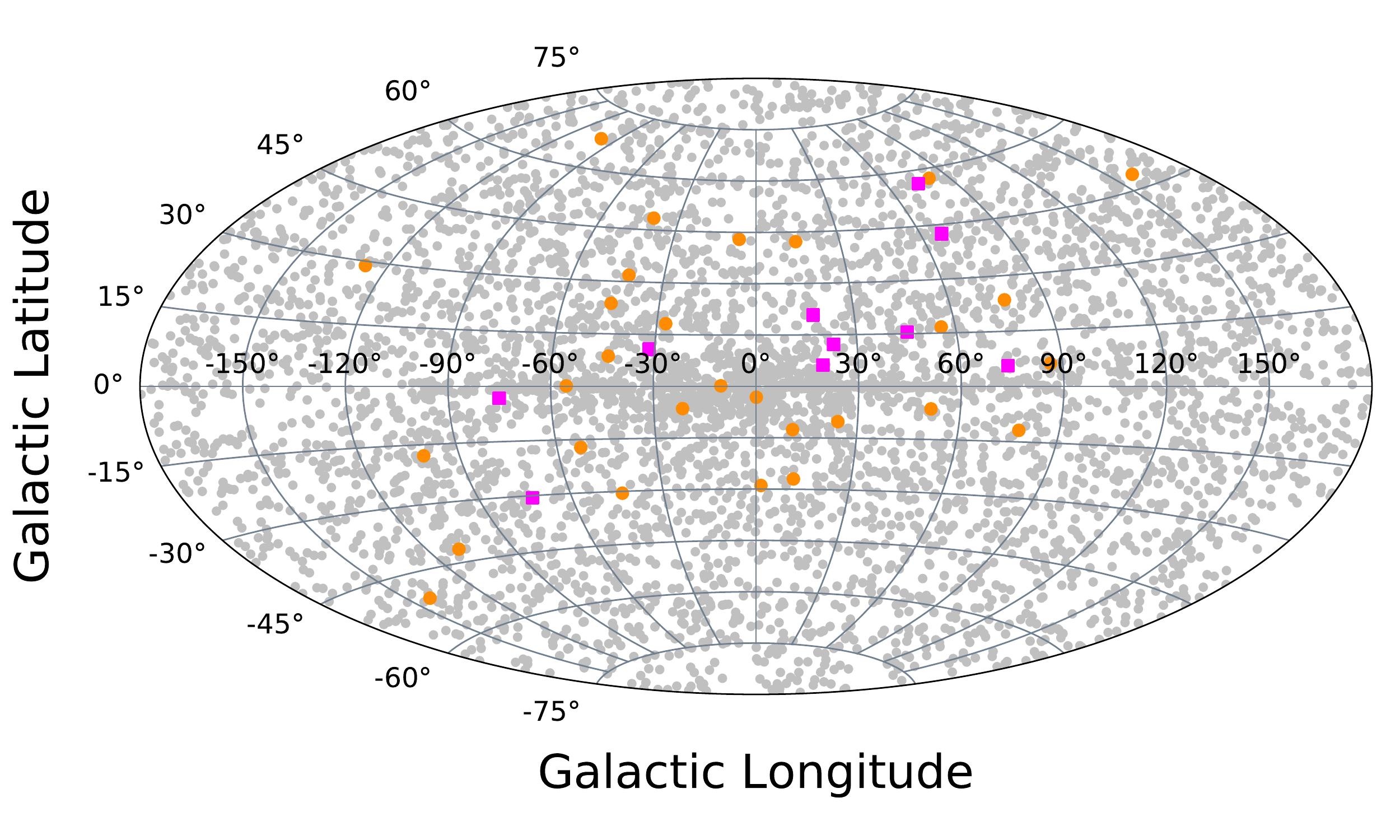}
   \caption{Galactic coordinates for 4FGL-DR1 sources (small grey dots), UID outliers (magenta squares) and identified sources passing the outlier selection (orange dots).}
  \label{fig:strange_lb}
\end{figure}

Examining the spectral and variability of indices 
for the outliers, they appear to be non-variable sources with a slightly curved spectrum and a spectral index close to 2.4. The position in galactic coordinates for the selected sources, as shown in Figure~\ref{fig:strange_lb}, appears to be limited to a diagonal band crossing the galactic centre. Similar behaviour is found for identified sources passing the same outlier selection. Perhaps the source's outlier nature is related to analysis artefacts or non-perfect diffuse emission modelling rather than to the intrinsic gamma-ray emission characteristics.

\section{Conclusions}
\label{sec:conclusions}

We characterised the sources detected by the \emph{Fermi}-LAT and listed in the 4FGL DR1 and DR2 catalogues using their gamma-ray spectral and light curve properties in an ensemble of different artificial neural network models trained for a three-category classification between BL Lac type objects, FSRQ and pulsars. DR1 and DR2 results are based on similar but independent ANN.

We found that many BLLs and FSRQs are classified as blazar-like even when they are not strongly associated with any of the two subclasses. The BCUs showed similar behaviour with more than 90 \% of the sample correctly characterised as blazars. About 80 \% of the globular clusters, whose gamma-ray emission is related to millisecond pulsars, are correctly classified as PSR.
The other subclasses of associated sources populated different regions of the $\alpha_{BLL}$, $\alpha_{FSRQ}$ and $\alpha_{PSR}$ phase space depending on their type but with substantial overlap between different classes, therefore, it is not possible to distinguish the different subclasses only using the classification presented in this work. All the sources belonging to classes related to active galaxies are closely classified as blazar-like.
SPP classification pointed to a possible difference with the potentially associated populations of SNR and PWN.

Starting from a relatively unbalanced population among the three classification categories, the ANN produced approximate equipartition of the unassociated sources, resulting in a number of potential pulsars exceeding the already known ones. In addition, the candidate pulsars follow the expected pattern in galactic latitude.  Also candidate blazars  conform to expectations in terms of galactic latitude distribution but with a difference between BLL-like and FSRQ-like UID of unclear origin.

Searching for outliers among the 4FGL unassociated sources, we found an unusual fraction of high significance blazar-like UID with intermediate blazar likelihood.
We selected 10 UID sources showing the same peculiarity both in DR1 and DR2 classifications as potentially interesting for follow-up studies, even if the galactic coordinates 
distribution of the selected sources may hint at an artefact related to the galactic diffuse emission model.

\section*{Acknowledgements}

Support for science analysis during the operation phase is gratefully acknowledged from the Fermi-LAT collaboration for making the Fermi-LAT results available in such a useful form,  the Department of Physics and Geology of the University of Perugia – Italy, National Institute for Astrophysics (INAF) Rome – Italy, and National Institute for Nuclear Physics (INFN) Rome – Italy.
The authors would like to thank  Dr. J. Tingey (University College London) for reviewing the manuscript and the anonymous referee for discussion and suggestions leading to the improvement of this work.

\section*{Data Availability}

The data underlying this article are available in the article and in its online supplementary material.








\appendix

\section{Source Classes Acronyms}
\label{app:acronyms}
	\begin{tabular}{ll} 

            AGN   & other non-blazar active galaxy\\
            BCU   & active galaxy of uncertain type\\
            BIN   & binary\\
            BLL   & BL Lac type of blazar\\
            CSS   & compact steep spectrum quasar\\
            FSRQ  & FSRQ type of blazar\\
            GAL   & normal galaxy (or part)\\
            GLC   & globular cluster\\
            HMB   & high-mass binary\\
            LMB   & low-mass binary\\
            MC    & molecular cloud\\
            NLSY1 & narrow line Seyfert 1\\
            NOV   & nova\\
            PSR   & pulsar\\
            PWN   & pulsar wind nebula\\
            RDG   & radio galaxy\\
            SBG   & starburst galaxy\\
            SEY   & Seyfert galaxy\\
            SFR   & star-forming region\\
            SNR   & supernova remnant\\
            SPP   & potential association with SNR or PWN\\
            SSRQ  & soft spectrum radio quasar\\
            UID   & unassociated \\
            UNK   & unknown\\
    \end{tabular}

\section{Supporting Material}

The full tables of the DR1 and DR2 classification results are available online on the MNRAS web site.


\bsp	
\label{lastpage}

\end{document}